# Precise measurement of parity violation in polarized muon decay

J. F. Bueno,[1,*] R. Bayes,[2,†] Yu. I. Davydov,[2,‡] P. Depommier,[3] W. Faszer,[2] C. A. Gagliardi,[4] A. Gaponenko,[5,§] D. R. Gill,[2] A. Grossheim,[2] P. Gumplinger,[2] M. D. Hasinoff,[1] R. S. Henderson,[2] A. Hillairet,[2,∥] J. Hu,[2,¶] D. D. Koetke,[6] R. P. MacDonald,[5] G. M. Marshall,[2] E. L. Mathie,[7] R. E. Mischke,[2] K. Olchanski,[2] A. Olin,[2,**] R. Openshaw,[2] J.-M. Poutissou,[2] R. Poutissou,[2] V. Selivanov,[8] G. Sheffer,[2] B. Shin,[2,††] T. D. S. Stanislaus,[6] R. Tacik,[7] and R. E. Tribble[4]

(TWIST Collaboration)

[1]*University of British Columbia, Vancouver, British Columbia, V6T 1Z1, Canada*
[2]*TRIUMF, Vancouver, British Columbia, V6T 2A3, Canada*
[3]*University of Montreal, Montreal, Quebec, H3C 3J7, Canada*
[4]*Texas A&M University, College Station, Texas 77843, USA*
[5]*University of Alberta, Edmonton, Alberta, T6G 2J1, Canada*
[6]*Valparaiso University, Valparaiso, Indiana 46383, USA*
[7]*University of Regina, Regina, Saskatchewan, S4S 0A2, Canada*
[8]*Kurchatov Institute, Moscow, 123182 Russia*
(Received 18 April 2011; published 9 August 2011)

We present a new high precision measurement of parity violation in the weak interaction, using polarized muon decay. The TWIST Collaboration has measured $P_\mu^\pi \xi$, where $P_\mu^\pi$ is the polarization of the muon in pion decay and $\xi$ describes the intrinsic asymmetry in muon decay. We find $P_\mu^\pi \xi = 1.000\,84 \pm 0.000\,29(\text{stat.})^{+0.001\,65}_{-0.000\,63}(\text{syst.})$, in good agreement with the standard model prediction of $P_\mu^\pi = \xi = 1$. Our result is a factor of 7 more precise than the pre-TWIST value, setting new limits in left-right symmetric electroweak extensions to the standard model.

     

## I. INTRODUCTION

Of the fundamental interactions, only the weak interaction is not symmetric under the parity transformation. The standard model (SM) of charged electroweak interactions assumes maximal parity violation based on empirical evidence. However, SM extensions with additional weak couplings can restore parity conservation at higher energies, and precision studies of muon decay are a method of probing such interactions. Within the SM, positive muons decay to a positron and two neutrinos via left-handed couplings to the $W$-boson. For muons that originate from pion decay, the forward-backward asymmetry of the positron from muon decay can be described by $P_\mu^\pi \xi$. The parameter $P_\mu^\pi$ is the polarization of the muon with respect to its momentum vector at the instant of pion decay, and $\xi$ describes the intrinsic asymmetry that is characteristic of parity violation.

*jbueno@triumf.ca
†Present address: University of Glasgow, Glasgow, United Kingdom
‡Present address: JINR, Dubna, Russia
§Present address: LBNL, Berkeley, California, USA
∥Present address: University of Victoria, Victoria, British Columbia, Canada
¶Present address: AECL, Mississauga, Ontario, Canada
**Also at University of Victoria, Victoria, British Columbia, Canada
††Also at University of Saskatchewan, Saskatoon, Saskatchewan, Canada

In a treatment [1] more general than the SM, where muon decay proceeds via an interaction that is local, derivative-free, Lorentz-invariant, and lepton-number-conserving, the matrix element $M$ can be written in terms of chiral amplitudes as

$$M \sim \sum_{\substack{\gamma=S,V,T \\ \epsilon,\mu=L,R \\ (n,m)}} g^\gamma_{\epsilon\mu} \langle \bar{e}_\epsilon | \Gamma^\gamma | (\nu_e)_n \rangle \langle (\bar\nu_\mu)_m | \Gamma_\gamma | \mu_\mu \rangle, \quad (1)$$

where the indices $\epsilon$ and $\mu$ label the electron and muon chiralities, $n$ and $m$ label the chirality of the neutrinos, $g^\gamma_{\epsilon\mu}$ are complex amplitudes, and $\Gamma^\gamma$ are the possible interactions (scalar/pseudoscalar (S), vector/axial-vector (V), and tensor (T)). For the SM, $g^V_{LL} = 1$ and the other amplitudes are zero, corresponding to a vector minus axial-vector interaction $(V - A)$.

For a positive muon, the most general differential decay rate can be written in terms of the four muon decay parameters $(\rho, \delta, \xi, \eta)$ [2] as

$$\frac{d^2\Gamma}{dx d\cos\theta_s} = \frac{m_\mu}{2\pi^3} W^4_{e\mu} G^2_F \sqrt{x^2 - x_0^2} \{F_{IS}(x) + P_\mu \xi \cos\theta_s F_{AS}(x)\}, \quad (2)$$

$$F_{IS}(x) = x(1-x) + \frac{2}{9}\rho(4x^2 - 3x - x_0^2)$$
$$+ \eta x_0(1-x) + F_{IS}^{RC}(x), \qquad (3)$$

$$F_{AS}(x) = \frac{1}{3}\sqrt{x^2 - x_0^2}[1 - x + \frac{2}{3}\delta(4x - 3 + (\sqrt{1-x_0^2}-1))]$$
$$+ F_{AS}^{RC}(x), \qquad (4)$$

where $G_F$ is the Fermi coupling constant, $W_{e\mu}$ is the electron's maximum energy (52.83 MeV), $x = E_e/W_{e\mu}$, $x_0$ is the minimum positron energy (0.511 MeV) divided by $W_{e\mu}$, $\theta_s$ is the angle between the muon spin and positron momentum vector, and $P_\mu = |\vec{P}_\mu|$ is the degree of muon polarization at the time of decay. $F_{IS}$ is the isotropic contribution, while $F_{AS}$ is the asymmetric contribution. The rate has been summed over the unobserved decay positron polarization. Since $P_\mu\xi$ only appears as a product in Eq. (2), we cannot measure $P_\mu$ and $\xi$ as separate quantities. The superscripts $RC$ in Eqs. (3) and (4) refer to radiative corrections. In the SM, the muon decay parameters are $\rho = \delta = 3/4$, $\eta = 0$, $\xi = 1$, and the muon polarization is $P_\mu^\pi = 1$.

Differences of the muon decay parameters from the SM values can be interpreted in terms of models of new physics. In left-right symmetric (LRS) electroweak models [3], a vector plus axial vector $(V + A)$ interaction is introduced that couples leptons of right-handed chirality, restoring parity conservation at high energies. The vector bosons of the $(V - A)$ and $(V + A)$ interactions are then $W_L$ and $W_R$, related to the mass eigenstates $W_1$ and $W_2$ by

$$W_L = W_1\cos\zeta + W_2\sin\zeta,$$
$$W_R = e^{i\omega}(-W_1\sin\zeta + W_2\cos\zeta), \qquad (5)$$

where $\zeta$ is a mixing angle and $\omega$ is a $CP$-violating phase. $W_R$ is much heavier than $W_L$, and apparent parity violation at low energies is a result of the mass difference. The left- and right-handed interactions have separate coupling constants, $g_L$ and $g_R$. The model contains right-handed neutrinos. If these neutrinos are light on the scale of the muon mass, and there is no mixing in the leptonic sector, then $\delta$ and $\eta$ retain their SM values but $P_\mu$, $\xi$, and $\rho$ have the relationships [4]

$$P_\mu \simeq 1 - 2t_\theta^2 - 2\zeta_g^2 - 4t_\theta\zeta_g\cos(\alpha + \omega), \qquad (6)$$

$$\xi \simeq 1 - 2(t^2 + \zeta_g^2), \qquad (7)$$

$$\rho \simeq \frac{3}{4}(1 - 2\zeta_g^2), \qquad (8)$$

where

$$t = \frac{g_R^2 m_1^2}{g_L^2 m_2^2}, \qquad (9)$$

$$t_\theta = \frac{g_R^2 m_1^2}{g_L^2 m_2^2}\frac{|V_{ud}^R|}{|V_{ud}^L|}, \qquad (10)$$

$$\zeta_g = \frac{g_R}{g_L}\zeta, \qquad (11)$$

$\alpha$ is a $CP$ violating phase in the right-handed Cabibbo-Kobayashi-Maskawa (CKM) matrix, and $V_{ud}^{L,R}$ are the elements of the left- and right-handed CKM matrices. This is the generalized (or nonmanifest) LRS model, in which $P_\mu\xi$ and $\rho$ allow limits to be set on the mass ratio, $t$, and the mixing angle, $\zeta_g$.

There are specific cases of LRS models that make further assumptions. In manifest LRS models [5], the left- and right-handed CKM matrices are assumed to be the same, $g_L = g_R$, and $\omega = 0$ so that $t_\theta = t$ and $\alpha = 0$. Equations (6) and (7) then reduce to

$$P_\mu\xi \approx 1 - 4t^2 - 4\zeta^2 - 4t\zeta, \qquad (12)$$

so that

$$\zeta = \frac{1}{2}(-t \pm \sqrt{1 - P_\mu\xi - 3t^2}). \qquad (13)$$

The goal of TWIST was to search for new physics that could be revealed by measurements of the muon decay parameters $\rho$, $\delta$, and $P_\mu^\pi\xi$ with an order of magnitude higher precision than previously achieved. Earlier TWIST results have been presented in [6–10]. An overview of the final TWIST measurements has been reported in [11]. This article, the first of two presenting a detailed description of these final TWIST measurements, describes the measurement of $P_\mu^\pi\xi$. We discuss the apparatus, data-taking procedures, and analysis techniques, with a particular focus on the improvements relevant to $P_\mu^\pi\xi$ that were made for this measurement. We then provide a detailed discussion of the systematic uncertainties associated with the determination of the polarization of the muon at the time of decay. These dominate the total uncertainty in $P_\mu^\pi\xi$ and are not relevant for the measurements of $\rho$ and $\delta$. In contrast, the systematic uncertainties associated with the measurement of the decay spectrum shape dominate the uncertainties in $\rho$ and $\delta$, while making only small contributions to the uncertainty in $P_\mu^\pi\xi$. Detailed discussion of the final $\rho$ and $\delta$ measurements, including the evaluation of the latter uncertainties, will be provided in a second publication [12].

## II. APPARATUS

The TRIUMF cyclotron in Vancouver, Canada, produces a proton beam of over $100\mu A$ with kinetic energy of 500 MeV and a time structure consisting of bunches of 2–4 ns width separated by 43 ns. In the TWIST experiment the protons interacted with a graphite production target of thickness 1.0 cm to produce pions and other secondary particles. A positive pion at rest decays into a muon and neutrino, each of momentum 29.79 MeV/$c$. In this frame, muons have 100% negative helicity in the SM, except for the radiative decay mode of the pion and the finite mass of the muon neutrino, which change the helicity at the level of $10^{-5}$ (negligible for our purposes). The M13 beam line (Fig. 1) was tuned to select particles at 29.60 MeV/$c$, which included muons from pion decay at rest in a thin outer layer of the graphite target. These are known as surface muons since they originate near the surface of the graphite target [13]. Except for a small correction due to muon scattering while exiting the surface layer, the near-perfect helicity was maintained by the beam transport.

The M13 dipole magnets B1 and B2 selected the average beam momentum. The beam optics were controlled using three vertically focusing (Q1, Q4, Q7) and four horizontally focusing quadrupole magnets (Q2, Q3, Q5, Q6). A series of apertures was used to select the beam line acceptance and a momentum resolution of 0.7% (FWHM), as measured by the width of the high-momentum kinematic edge of the surface muon distribution. The muon rate at the end of the beam line was between 2000 s$^{-1}$ and 5000 s$^{-1}$. The elements of M13 and the proton beam line were continuously monitored with a system that recorded information such as currents, voltages, and temperatures. The system also recorded the B1 and B2 fields using NMR probes.

Near the exit of the final quadrupole of the M13 channel, a pair of orthogonal time expansion chambers (TECs) could be inserted into the beam line vacuum system to measure the position and angle of individual muon trajectories [14]. The TECs were filled with dimethyl ether gas at a low pressure of 8 kPa, in order to keep the mass of materials as low as possible. This minimized multiple Coulomb scattering (MCS), which degraded the TEC resolution and also the muon polarization. Despite these efforts, MCS meant that the TECs could not remain in place during normal data acquisition. The modules were typically inserted on a weekly basis, at the beginning and end of each data set (during which the beam line settings were not altered), for approximately 1 hour of beam measurements. The characteristics of a typical nominal beam are shown in Fig. 2.

Each TEC had 24 sense wires, but as a result of the low operating pressure, on average only 15 wires produced signals for each muon. The TEC sense planes lost efficiency with time, and in our previous $P_\mu^\pi \xi$ measurement [8] the average number of points per track dropped to just

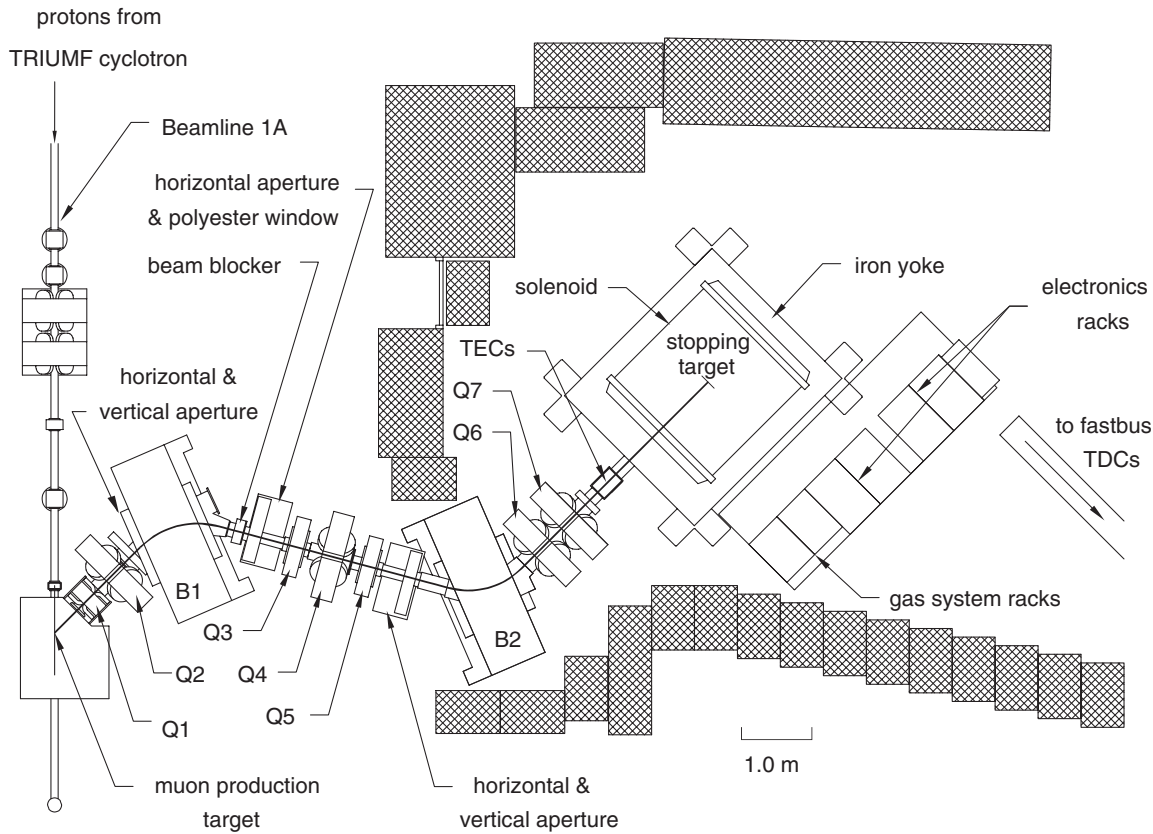

FIG. 1. Proton and M13 beam lines at TRIUMF. Surface muons were selected from the muon production target and transported to a muon stopping target at the center of the TWIST detector. The properties of the muon beam were measured by a pair of time expansion chambers (TECs) near the entrance of the detector.

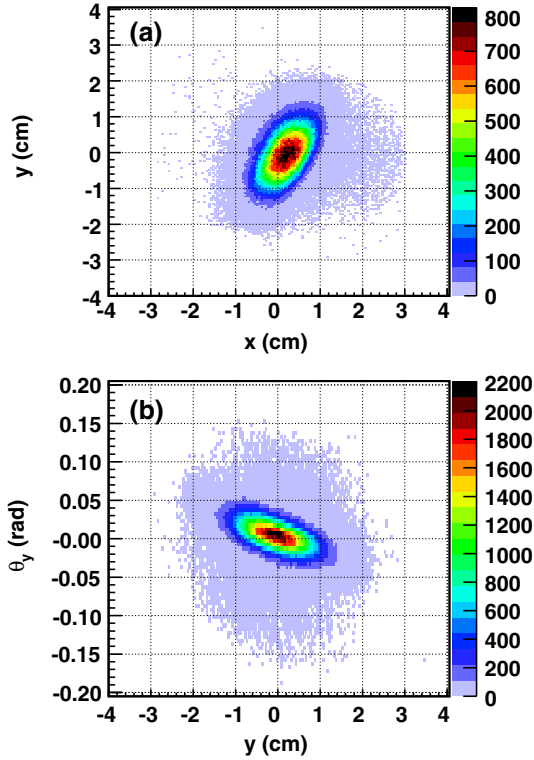

FIG. 2 (color online). Measurement of the nominal muon beam (a) position and (b) convergence using the TECs. The $x$ and $y$ coordinates are orthogonal to the muon beam direction. The position $x = y = 0$ corresponds to the symmetry axis of the solenoidal magnetic field. The correlation between $y$ and $\theta_y$ demonstrates that the muon beam is converging in this plane.

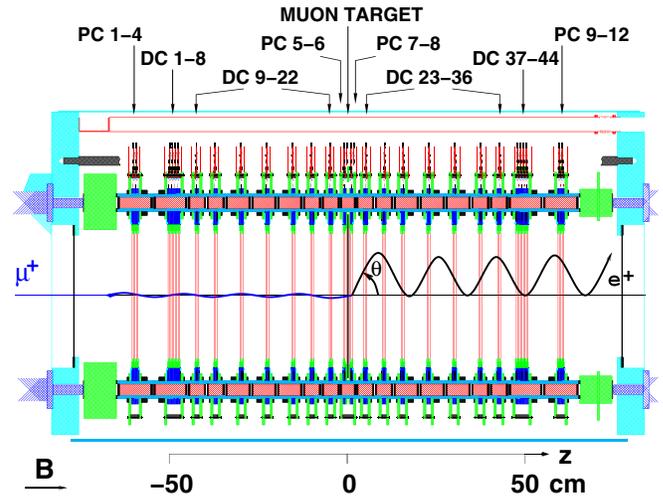

FIG. 3 (color online). A cross section of the TWIST detector array, including an example event. Proportional chambers (PCs) provided timing information and drift chambers (DCs) determined the position of particles. The angle between the decay positron and the beam direction is $\theta = \pi - \theta_s$.

four. This aging effect was eliminated by regularly changing the TEC sense planes after several hours of exposure to the beam.

Muons passed from the final beam line quadrupole through an extension of the beam line vacuum system into the TWIST spectrometer, consisting of a superconducting solenoid in an iron yoke with an array of high precision position-sensitive planar detectors within the room-temperature bore of the solenoid. The vacuum extension was part of an upstream beam package, which also included a Mylar vacuum window, a variable density gas degrader, an adjustable foil degrader, a thin muon scintillator, and a scintillator to detect decay positrons. The variable density gas degrader volume was a 20 cm length of $CO_2$ and He at atmospheric pressure contained between the 100 $\mu$m vacuum window on one side and a 6 $\mu$m Mylar membrane on the other. Natural variations in external pressure and temperature caused the gas density within the TWIST spectrometer to change, but the muon stopping position was kept constant by a feedback loop that changed the ratio of $CO_2$ and He and thus the density of the gas degrader. The foil degrader could be used to place up to 0.1 cm of Mylar in the path of the muons, but it was typically set for zero thickness. The muon scintillator was a plastic disc of radius 3.0 cm and thickness 239 $\mu$m, providing a trigger to the data acquisition signaling the arrival of a muon in the spectrometer. It was surrounded by an annular scintillator (inner radius 3.0 cm, outer radius 18.5 cm) that detected positrons decaying in the upstream direction. After the upstream beam package, the muons entered the TWIST detector array (Fig. 3) and about 80% stopped in a thin metal foil at its center. A set of scintillators was added at the downstream end of the detector to provide reliable timing information for all downstream decay positrons. For one week of data taking, a downstream beam package was added that mirrored the upstream beam package, except that the muon and positron scintillators were inactive; its purpose was to study the effect of backscattered positrons.

The detector array, or stack, was located centrally in the bore of the superconducting solenoid, which itself was within a custom iron yoke to produce a highly uniform magnetic field of 2 T nominal strength at its center. The yoke also restricted the range of the fringe field to approximately 0.1 T at the position where the TECs were inserted. The solenoid was operated in persistent mode. The observed decay of the central field strength ($< 10 \mu$T per month, monitored by two NMR probes) was taken into account during the analysis.

The detector stack comprised 44 planar multiwire drift chambers (DCs) and 12 multiwire proportional chambers (PCs), symmetrically positioned about the stopping target [15]. The chambers were very thin ($\approx 1 \times 10^{-4}$ radiation lengths) in order to minimize positron energy loss and MCS. Each drift chamber was filled with dimethyl ether gas, which has a slow drift velocity and small Lorentz angle. Drift times could be up to $\sim 1 \mu$s, and the position resolution was between 50 and 100 $\mu$m (rms) [16]. The drift chambers contained 80 wires of diameter 15 $\mu$m at a pitch of 0.4 cm. The cathodes were aluminized Mylar foils of thickness 6.35 $\mu$m, separated by 0.4 cm. Proportional chambers measured timing and energy deposited; they had the same construction as the drift chambers, except the wire pitch was 0.2 cm and the gas was a mixture of $CF_4$ and isobutane, which has high drift velocity. The PC time resolution was typically <20 ns.

The DCs and PCs were surrounded by helium gas, with 3% molecular concentration of nitrogen added to prevent high voltage sparking. A pressure control system kept the cathode foils flat by maintaining a differential pressure of less than 0.3 Pa between the helium and chamber gases.

Drift chambers were assembled into modules of pairs of orthogonal planes, at ±45° with respect to the horizontal, with a common cathode foil. Figure 3 shows that DCs 1–8 and 37–44 were each single modules consisting of four adjacent pairs of planes. The modules making up DCs 9–22 and 23–36 consisted of only two planes each. The z-spacing between the two-plane modules originally alternated between 5.2 cm and 7.2 cm, but this periodicity was found to cause an ambiguity in determining the radius of the positron's helical trajectory. For the current measurement, the spacing was changed to 5.2, 5.2, 7.2, 7.2, 7.2, 5.2, and 5.2 cm.

The nominal operating voltage was 1950 V for the DCs and 2050 V for the PCs, corresponding to an efficiency for positron detection of 99.8% for each chamber. However, the voltage on the PCs immediately before the target (PCs 5 and 6 in Fig. 3) was reduced to 1600 V. As a result, these PCs were inefficient for positrons, but their pulse widths were unsaturated and had greater sensitivity to muon energy loss, allowing us to discriminate against muons that stopped in the gas immediately before the metal stopping target.

Signals from all chambers were read with LeCroy 1877 fastbus time-to-digital converters (TDCs) that recorded rise time and time-over-threshold. The TDCs had integral and differential nonlinearities of <25 ppm (full range) and <0.1 ns [17], respectively, which are negligible for our purposes.

Muons were stopped in a single thin target foil. Data were taken with an Al target of 71.0 $\mu$m thickness that had been used previously [8,9], and a new Ag target of 30.9 $\mu$m thickness. Both targets were of purity exceeding 99.999%. The target foil served as a common cathode foil for PC 6 and PC 7.

## III. MUON BEAM SELECTION AND OPTIMIZATION

The beam had proton, positron, pion, and muon components with central momentum selected by B1 and momentum spread determined by subsequent apertures in the beam. The very low energy protons lost a significant fraction of their momentum while passing through the 3 $\mu$m polyester membrane between B1 and Q3 (see Fig. 1) and thus did not survive a second momentum selection by B2. The polyester window was used to prevent neutral radioactive products emitted by the graphite pion production target from reaching the TWIST detector, while reducing the muon polarization by only ~$10^{-6}$ due to MCS. The beam positrons originated from muon decays within the production target and surrounding materials, as well as from gamma conversions in the region of the production target. The positron rate was several times higher than the muon rate, but it did not affect the data quality since beam positrons passed through the entire detector array and were easily identified by the analysis software. Only a small fraction of pions survived to the detector because their decay length at this momentum was only 1.6 m; those that entered the detector did not reach the stopping target due to their greater energy loss compared to muons.

In the TWIST measurement of muon decay, muon polarization, $P_\mu$, is measured with respect to the detector's z-axis (see Fig. 3). In this paper, depolarization refers to any reduction in polarization with respect to the z-axis. At the time of muon production, the polarization direction is opposite to the muon momentum, so that any real beam with nonzero mean square transverse momentum (i.e., nonzero divergence) has a lower polarization along a geometric axis. The amount of depolarization can be estimated at the position of the TECs because they provide a measure of the transverse momentum components of the beam.

The time structure of the TRIUMF proton beam allowed selection of muons with high polarization (surface muons). Figure 4 shows the trigger time relative to a probe in the proton beam line and the relative polarization of the beam, as determined by the forward-backward asymmetry that was observed in the detector. Because of the 10 m length of the M13 beam line, pions at the selected beam momentum (29.60 MeV/$c$) arrived approximately 170 ns after production by a proton in the graphite target. Muons left the target with a time distribution governed by the 26 ns lifetime of the pion at rest in the target, and then took about 130 ns to reach the detector. Pions thus arrived almost one full accelerator repetition period after the earliest surface muons. A contamination of cloud muons was also present; these originated from pions decaying in flight while moving between the production target and the first dipole (B1). They arrived at the same time as the earliest surface muons, but they had a small, opposite, and poorly determined polarization. The difference in polarization for the cloud muon time region is clear from the forward-backward asymmetry. A selection of particles arriving between 10 and 30 ns after the earliest surface muons is uncontaminated by either cloud muons or pions, as evidenced by the consistent large asymmetry.

As muons travel from the end of the beam line into the field of the solenoid, polarization is further reduced as

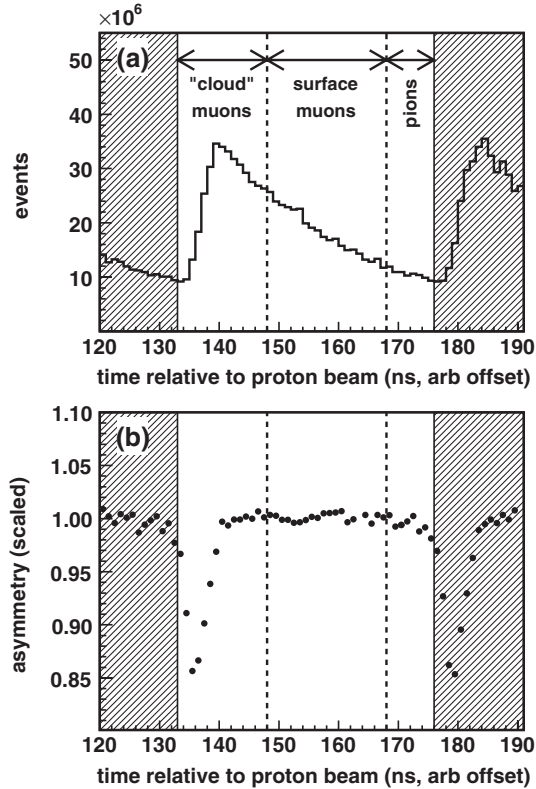

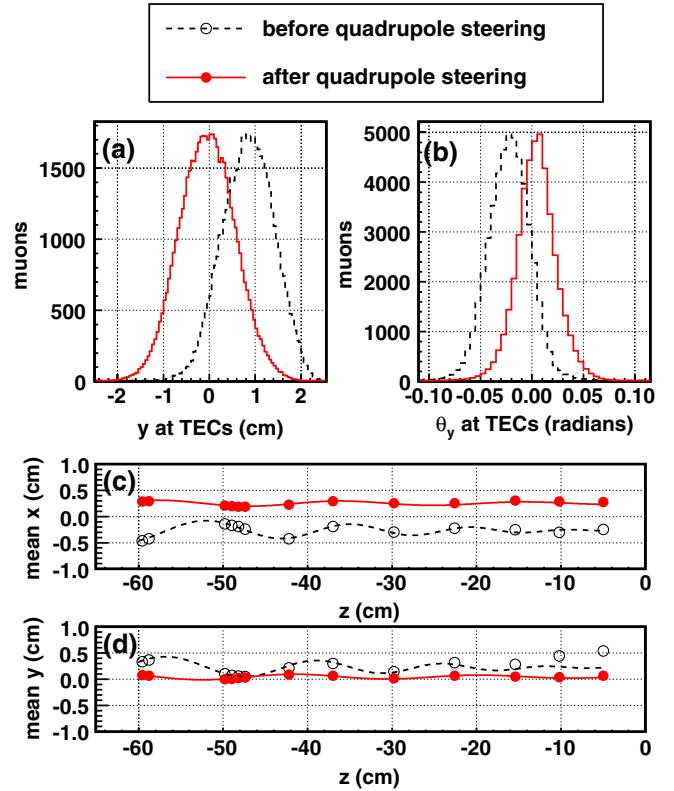

FIG. 4. (a) Number of events as a function of time relative to the proton beam. The spectrum shape corresponds to the pion lifetime. (b) The resulting forward-backward asymmetry of the decay positrons.

FIG. 5 (color online). Muon beam improvements. (a) The beam was steered onto the solenoid symmetry axis ($x = y = 0$), and (b) a vertical average angle of $\approx 20$ mrad was removed. This reduced the transverse momentum of the beam inside the spectrometer, as measured by the mean position of the beam at each pair of planes (c) and (d).

the fringe field increases transverse momentum and the longitudinal magnetic field increases from $\approx 0.1$ T to $\approx 2.0$ T. The loss of polarization is minimized by directing the muon beam close to the solenoid's symmetry axis, with small transverse momentum. Such a beam also has the smallest uncertainty in final polarization when estimated by a simulation of spin evolution in the fringe field.

For our previous $P_\mu^\pi \xi$ measurement, only steering in the horizontal direction using the B2 dipole was available to align the muon beam with the detector axis. The resulting beam was not optimally aligned with the solenoidal magnetic field; the vertical position was $\approx 1$ cm above the solenoid's symmetry axis [Fig. 5(a)], and the beam had a net angle of $\approx 20$ mrad [Fig. 5(b)]. For the measurements reported here, quadrupole steering was added by installing current sources that could be used to increase the field in individual pole pieces of three of the M13 quadrupoles (Q4, Q6, and Q7). When an asymmetric current was applied to the poles, a dipole field was added that shifted the zero field location away from the quadrupole center. As a result, the muon beam was both focused and steered [18]. The average position and angle of the muon beam at the TECs had a linear dependence on the asymmetric current applied to the quadrupoles and the field strength of the B2 dipole. The linear sensitivity of the beam properties at the

TECs to each of the four steering elements was measured, and then the elements were combined to set the desired average values of $x$, $y$, $\theta_x$, $\theta_y$. For nominal operation the beam was centered on the solenoid's symmetry axis ($x = y = 0$), with a net average angle of a few mrad. The resulting beam had reduced transverse momentum, as measured by the drift chambers [Figs. 5(c) and 5(d)].

## IV. DATA SETS

Data used for this analysis (Table I) were acquired from October 2006 to December 2006 (Ag target), and from May 2007 to August 2007 (Al target). Each data set was approximately one week in duration, over which time the running conditions were unchanged. The four nominal data sets had an initial mean muon momentum of 29.60 MeV/$c$, a stopping distribution with the Bragg peak near the center of the target, a magnetic field of central strength 2.0 T, and no downstream beam package in place. Set 68 was obtained with the stopping distribution moved upstream, which proved to be an important diagnostic for positron energy losses in the stopping target. The central field strength was decreased (increased) by 2% for set 70 (71), changing the radius of the positron helices. Sets 70 and 71 were then a consistency check that our measurements are unaffected by the transverse scale of the helices. For set 83 the downstream beam package was inserted; this allowed us to validate the simulation of back-scattered positrons. Sets 91, 92, and 93 were taken at lower muon beam momenta to select muons that passed through more production target material, and therefore experienced more MCS, to confirm that depolarization via MCS was understood. The reduction in momentum was limited by the need for muons to reach the stopping target.

TABLE I. Data sets used during the analysis, in the order they were acquired. The data sets numbered 68–76 (80–93) used an Ag (Al) stopping target. Set 89 used a special larger-radius Al target of lower purity.

| Data set | Description | Events ($\times 10^6$) | |
|---|---|---|---|
| | | Before cuts | Final spectrum |
| 68 | Bragg peak 1/3 into target | 741 | 32 |
| 70 | Central field at 1.96 T | 952 | 50 |
| 71 | Central field at 2.04 T | 879 | 45 |
| 72 | TECs in, nominal beam | 926 | 49 |
| 73 | Muons stopped at detector entrance | 1113 | ⋯ |
| 74 | Nominal | 580 | 32 |
| 75 | Nominal | 834 | 49 |
| 76 | Off-axis beam | 685 | 39 |
| 80 | Muons stopped at detector entrance | 363 | ⋯ |
| 83 | Downstream beam package in place | 943 | 49 |
| 84 | Nominal | 1029 | 43 |
| 86 | Off-axis beam | 1099 | 58 |
| 87 | Nominal | 854 | 45 |
| 91 | Lower momentum I ($p = 28.75$ MeV/$c$) | 225 | 11 |
| 92 | Lower momentum II ($p = 28.85$ MeV/$c$) | 322 | 15 |
| 93 | Lower momentum III ($p = 28.85$ MeV/$c$) | 503 | 26 |
| 89 | Muons stopped at detector entrance | 708 | ⋯ |
| | | Total | 543 |

The data sets with the off-axis beam (76 and 86) were used to evaluate the largest $P_\mu^\pi \xi$ systematic uncertainty (see Sec. VII A). In these data sets, the transverse momentum of the beam was increased to deliberately lower the final polarization of the muons. For the Ag target (set 76), an average angle $\langle \theta_y \rangle$ of $\approx 30$ mrad was introduced at the TECs, and for the Al target (set 86), the beam was steered so that $\langle x \rangle$ was $\approx -1$ cm, and $\langle \theta_x \rangle$ was $\approx -10$ mrad. Set 72 was acquired with the TECs in place, to measure the long-term stability of the muon beam and also the aging of the TEC sense planes. This data set was also used as part of the systematic uncertainty evaluation since it increased the width of the angle distributions in the beam, allowing the muons to sample a broader region of the solenoid's fringe field.

For sets 73, 80, and 89, the muons were stopped immediately after the trigger scintillator, so that a single decay positron was reconstructed independently in each half of the detector. These data were used to investigate the detector response, not to extract muon decay parameters.

## V. SIMULATION

A detailed simulation of the detector was implemented in GEANT 3.21 [19]. The simulation facilitated a blind measurement of the muon decay parameters (Sec. VI), allowed study and optimization of the positron reconstruction algorithms, and predicted the final polarization of the muons by following the evolution of their momenta and spins to the stopping location.

Simulated events started with the generation of muons and beam positrons at the position of the TECs, with rates that matched the real data-taking conditions. The initial spin vector of each muon was antiparallel to its momentum direction. Pions, protons, and nonsurface muons were not simulated since they were removed from the real data, as described in Sec. III.

Initial muon particle trajectories were generated based on TEC measurements. The position was selected by dividing the $(x, y)$ distribution of the number of muons into 0.1 cm × 0.1 cm bins [Fig. 2(a)] and then applying an acceptance-rejection method. The angles were drawn from independent Gaussian distributions with a mean angle that matched the data measurement for the particular $(x, y)$ bin. The widths of the Gaussian angle distributions measured by the TECs were systematically too large due to MCS in the drift gas and windows. In order to correct for the MCS, the observed widths were multiplied by 0.64 in the $x$-module and 0.48 in the $y$-module. The multiplication factors were determined by simulating the TECs in GEANT and adjusting the factors until the distributions from the simulated TECs reproduced the data. The beam positrons were generated with the same algorithm as the muons, but the beam measurement came from upstream drift chamber data taken while the magnetic field was off.

The initial muon momenta were generated by the simulation according to a Gaussian distribution to represent the momentum acceptance of the beam line, but truncated above the maximum muon kinetic energy of 29.79 MeV/$c$. A dispersion effect was also included to account for an observed dependence of momentum on the $x$-position at the TECs. This was characterized by a linear relationship with $dp/dx = 0.17$ MeV/$c$ per cm, where the sensitivity was determined by analyzing runs with the TECs in place, fitting the average muon range as a function of the $x$-position as measured by the TECs and then tuning the simulation to match the data.

The simulation did not include any loss of muon polarization upstream of the TECs. However, as noted in Sec. III, the initial polarization of the beam was less than 1.0 due to the finite divergence of the beam. Also, a correction was applied to the final results to account for MCS in the graphite production target (Sec. VIII).

The spin vector, $\vec{s}$, was propagated through the solenoidal magnetic field, $\vec{B}$, using the Thomas-BMT equation [20],

$$\frac{d\vec{s}}{dt} = \frac{e}{mc} \vec{s} \times \left[ \left( \frac{g}{2} - 1 + \frac{1}{\gamma} \right) \vec{B} - \left( \frac{g}{2} - 1 \right) \frac{\gamma}{\gamma + 1} (\vec{\beta} \cdot \vec{B}) \vec{\beta} \right.$$
$$\left. - \left( \frac{g}{2} - \frac{\gamma}{\gamma + 1} \right) \vec{\beta} \times \vec{E} \right], \quad (14)$$

where $t$ is the time, $(e/m)$ is the charge to mass ratio, $c$ is the speed of light, $g$ is the Landé $g$-factor, $\gamma$ is the Lorentz factor, $\vec{\beta}$ is the velocity divided by $c$, and $\vec{E}$ is the electric field. The $E$ field term was not included in the simulation; the drift and proportional chambers had significant electric fields, but a field of $(-E)$ between the entrance cathode foil and the wire was followed by a symmetric $(+E)$ field between the wire and the exit cathode foil, and this effectively canceled the change in spin from the $E$ field term. The spin was propagated at each step using a Taylor expansion of Eq. (14), with the components of the spin vector renormalized to correct for numerical inaccuracies. For each GEANT particle transport step, smaller substeps were taken for the spin propagation, dependent on the magnitude of the magnetic field and muon velocity. The final spin tracking algorithm was demonstrated to be equivalent to a fourth-order Runge-Kutta numerical method for determining the absolute polarization at the level of $0.2 \times 10^{-4}$.

Depolarization from muon-electron scattering during deceleration was not included in the simulation; this is at the negligible level of $0.1 \times 10^{-4}$ for surface muons [21]. Depolarization from muonium formation was not included for two reasons. First, as muons approached the metal target, the muon velocity was typically much higher than the atomic electron velocity in the chamber materials, so that electron capture was heavily suppressed [22]. (Muons that slowed and stopped before reaching the metal stopping target were identified and removed via their energy loss in those chambers, as described in Sec. VI.) Second, muonium formation does not occur in the metal of the target foils and wires due to conduction electron screening.

There are three possible depolarizing mechanisms in our muon stopping targets: a hyperfine contact interaction between the muon spin and the conduction electron spins (Korringa relaxation [23]), an interaction with the magnetic field from the nuclear dipole moments of the host nuclei and impurities, and an interaction with paramagnetic impurities. For both targets Korringa relaxation is expected to be the dominant mechanism [24], and it has an exponential form,

$$P_\mu(t) = P_\mu(0) \exp(-\lambda t), \quad (15)$$

where $\lambda$ is a relaxation constant and $P_\mu(0)$ is the polarization at the time of muon thermalization, $t = 0$. The simulation applied Eq. (15) for muons stopped in the metal targets, with $\lambda$ determined from the data.

When a muon decayed, the simulation selected the positron energy and angle from a pregenerated spectrum that used hidden values for each of $\rho$, $\delta$, and $\xi$ that were randomly chosen from within a range of $\pm 0.01$ of the SM values. Radiative corrections of first order, leading-logarithmic and next-to-leading-logarithmic second order, and leading-logarithmic third order [25] were included. The positron was transported through the detector materials, and secondary particles (photons, positrons, and electrons) were also simulated. Default parameters were adopted in GEANT to simulate physics processes such as ionization energy loss, delta-electron production, bremsstrahlung, MCS, positron annihilation, Compton scattering, pair production, the photoelectric effect, and Bhabha scattering. Optional GEANT code was enabled to optimize the simulation of these processes within thin media.

Response of the drift and proportional chambers was simulated in detail, with the aim of accurately reproducing inefficiencies, bias, and resolution. As a particle passed through the chamber gas, ion clusters were randomly produced along the trajectory, with a mean cluster spacing that was tuned to match the data. The ion cluster position was converted into drift time using GARFIELD-generated [26] maps of space-time relationships. The time was then smeared to include effects of electronics and diffusion. Signals from wires that were dead in the data (one or two out of more than 3500) were deactivated in the simulation. Ionization from muon tracks caused the hit cell to be less efficient for subsequent positron hits. This inefficiency was quantified through studies of positron tracks intersecting cells with a muon hit and was implemented in the simulation by deadening the wire within 0.06 cm of the muon hit with a mean recovery time of 3.0 $\mu$s.

The output of the simulated wire chambers and scintillators was written to disk in the same binary format as the real data acquisition system. This enabled the simulated data to follow a nearly identical analysis procedure as the real data, including the application of alignment procedures, wire-by-wire time offset estimates, and chamber space-time relationship tuning [16]. By adding these imperfections to the simulation, systematic uncertainties were reduced and a better match was obtained, for example, for the detector resolution at the kinematic end point of the muon decay energy distribution.

Materials external to the tracking volume were included in the simulation, such as the beam pipe, the upstream beam package, and the glass frames for the chambers. The positrons could backscatter from these materials and reenter the tracking volume. The backscattered positrons introduced extra hits and resulted in a tracking inefficiency that the simulation reproduced.

## VI. ANALYSIS

The muon decay parameters were determined by applying an identical analysis to real data and the GEANT

simulation and then comparing the distribution of positron energy and angle (the spectrum) from data and simulation. The simulation's hidden decay parameters were revealed only at the end of the analysis. In this way we remained blind to the measured values of the decay parameters until all systematic uncertainties and corrections had been evaluated.

Reconstruction of data began by applying an offset to the time from every wire in the detector. This was calibrated on a wire-by-wire basis and accounted for variations in cable lengths and electronics. For the real data, crosstalk from induced signals within the DCs and the preamplifier cards was identified and removed. The DC signals for each particle were grouped into time windows of length 1.05 $\mu$s, which covers the great majority of drift times in a DC cell. For the purposes of particle identification and pattern recognition, a cluster position was calculated for signals from adjacent wires. A special treatment was necessary if two particles were separated by less than 1.05 $\mu$s [12].

The particle in each time window was tentatively identified: the pulse widths of signals in PCs 1–4 were used to distinguish between muons and positrons, since muons deposited more energy in the detector. Beam positrons were distinguished from decay positrons since the former typically passed through the entire detector, and the latter originated from the stopping target and only passed through half of the detector. The analysis code also looked for tracks that could be delta-electrons, positrons backscattering from material outside of the DC region, and trajectories that were apparently broken into two pieces due to discrete energy loss or large angle MCS. Events were then classified according to the particles observed and their time separation.

Time windows with decay positrons were assumed to contain a helix track, and an estimate was made of the radius, wavelength, and phase of the track using the position of the wires that were hit. A more precise fitting routine then converted the drift times in each DC to position and carried out a least squares fit that included continuous energy loss throughout the positron trajectory as well as the possibility of large angle scatters at each pair of planes. The helix-fitting code was reviewed and improved for this analysis [12], in particular, by a modification of the chamber space-time relationships based on data [16].

At this stage the essential parameters of each event had been evaluated, and selections were now applied to reject events where the simulation may have less accurately represented the real data. The important cuts are described explicitly below.

A time-of-flight cut was applied to the trigger particle in data to maximize the polarization, and the cut setting was tuned to make the observed asymmetry independent of time; the cut positions are shown in Fig. 4. There is no time-of-flight cut for the simulation since only 100% polarized muons were generated. Events with a single muon and at least one decay positron candidate were selected. The decay positron and muon had to be separated by more than 1.05 $\mu$s in order to avoid overlap of muon and positron ionization in the upstream half of the detector. Events were accepted only if the muon produced signals in PC 5 and PC 6, but not in PC 7. Muons that stopped in the gas or wires of PC 6 were removed based on the pulse widths in PC 5 and PC 6 [27]. Figure 6 shows the correlation between the pulse widths in the two chambers. Most muons that reached the target had positively correlated energy deposition in PC 5 and PC 6. This appears as a dark band through regions 1 and 2. When the muons failed to reach the target, they deposited a maximal amount of energy in PC 5 and progressively less energy in PC 6, and this appears as a second fainter vertical band through regions 2 and 3. Muons from region 1 were selected, corresponding to about 90% of the events.

Selections were applied to the momentum and angle of the decay positron track to define a fiducial region (Fig. 7). The maximum momentum cut ($p^{max} = 52.0$ MeV/$c$) avoided the region of the spectrum that was used in a momentum calibration procedure (described below). The longitudinal momentum cut ($|p_z^{min}| = 14.0$ MeV/$c$) avoided the region where the helix wavelength was difficult to determine. The requirement $|\cos\theta| < 0.96$ removed small angle tracks where the wavelength was poorly resolved, and $|\cos\theta| > 0.54$ eliminated large angle tracks with less reliable reconstruction due to MCS as the path length through the chambers became too large. The maximum transverse momentum cut ($p_t^{max} = 38.0$ MeV/$c$) retained only the positrons within the instrumented regions of the detector. The minimum transverse momentum cut ($p_t^{min} = 10.0$ MeV/$c$) removed tracks where the helix radius became comparable to the wire spacing. After all cuts, between 4% and 6% of original events remained in the spectrum (see Table I).

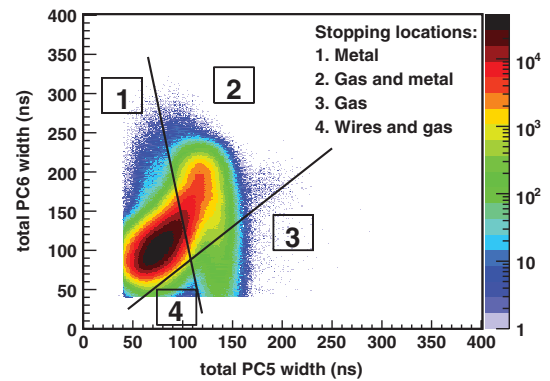

FIG. 6 (color). Distribution of pulse widths corresponding to ionization in the proportional chambers immediately before the target (PC 5 and PC 6). The muons in each quadrant predominantly stopped in the material indicated. The distribution was compared to simulation, confirming that region 1 contained an enriched sample of muons stopping in the metal target.

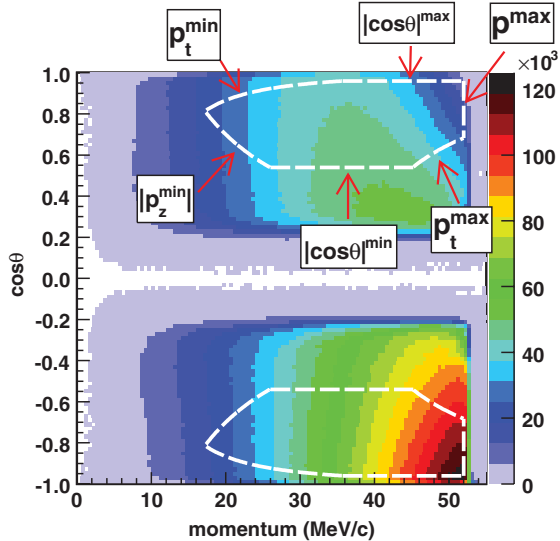

FIG. 7 (color). Reconstructed decay positron spectrum for the simulation of a single data set. The accepted events are inside the fiducial region defined by the total momentum ($p$), the transverse momentum ($p_t$), the longitudinal momentum ($p_z$), and the cosine of the angle ($\cos\theta$). The cuts are symmetric for upstream and downstream.

An iterative momentum calibration was applied while determining the difference in decay parameters between data and simulation. The calibration used the high-momentum portion of the spectrum that straddled the kinematic momentum end point (Fig. 8), where on average the simulation's reconstructed momentum was larger than in the data by $\sim 0.01$ MeV/$c$. The difference between the data and simulation was found for bins in $1/\cos\theta$, and these differences are fitted separately upstream and downstream to a slope ($a$) and an intercept ($b$) at $\cos\theta = 0$. We bring the data and simulation into agreement by applying two calibration models to the data spectrum. An

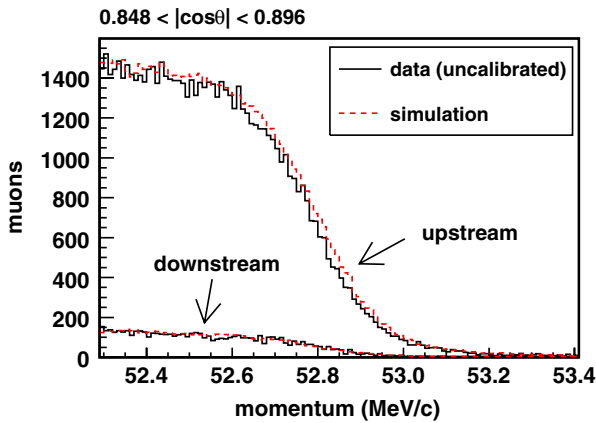

FIG. 8 (color online). Spectrum slice in $\cos\theta$ for a single data set and its simulation. The simulation's end point momentum is $\sim 0.01$ MeV/$c$ greater than data. The data spectra were calibrated so that the end point for all angle slices agreed with the simulation.

error in the magnetic field scale would mostly contribute a momentum scaling, so the first calibration model scaled the data using

$$p_{\text{corrected}} = \frac{p_{\text{reconstructed}}}{1 + \frac{1}{\sqrt{W_{e\mu}^2 - m_e^2}}\left(b - \frac{a}{|\cos\theta|}\right)}. \quad (16)$$

The other extreme would be a momentum shift, so the second model shifted the data to make its end point consistent with the simulation according to

$$p_{\text{corrected}} = p_{\text{reconstructed}} - \left(b - \frac{a}{|\cos\theta|}\right). \quad (17)$$

The two calibration models were applied independently between each data set and its simulation, and the central value for $P_\mu^\pi \xi$ was averaged over the two calibrations.

The differences $\Delta\rho$, $\Delta\delta$, and $\Delta P_\mu^\pi \xi$ in the three decay parameters between the unknown (blind) values of the simulation and the ones represented by the data were obtained by treating them as fit parameters in a chi-squared minimization over the two-dimensional selected momentum-angle ranges of the decay positron tracks [28]. This takes advantage of the linearity of Eq. (2) in a combination of these parameters. Bins were defined in intervals of 0.5 MeV/$c$ for momentum and 0.02 for $\cos\theta$, with 2442 bins falling completely within the selected regions. Normalization of the number of events in data and simulation was applied prior to the fit. While the hidden simulation values were not revealed until all systematic uncertainties and corrections had been established, the changes in the parameter differences could be used at an earlier stage to assess the influence of many systematic effects, some of which are described in the next section.

## VII. SYSTEMATIC UNCERTAINTIES

The precision of the $P_\mu^\pi \xi$ result is limited by systematic uncertainties, which were determined according to how well the simulation reproduced the real detector and physics processes. The systematic and statistical uncertainties are summarized in Table II. Most categories were determined simultaneously for all decay parameters, but those related to depolarization are unique to $P_\mu^\pi \xi$. The systematic uncertainties are grouped according to whether they were target independent or specific to the aluminum or the silver target.

### A. Depolarization in fringe field

The projection of muon spin vectors onto the detector axis, and thus muon polarization, evolved as beam particles passed into the high field region of the spectrometer. Assuming that the momentum and spin vectors were antiparallel at the location of the TECs, then $P_\mu \approx 0.9999$ at this location due to beam particles having a small range of angles [Fig. 2(b)]. At the entrance of the stopping target, the polarization was reduced to $\approx 0.9975$; the principle cause of this reduction in polarization was that transverse components of momentum and therefore of spin were generated as the muons encountered components of the fringe field perpendicular to their momentum vectors (see Fig. 9). The uncertainties in accounting for this depolarization will now be described; they depend on the quality of the TEC measurements and the accuracy of the magnetic field map in the region of the solenoid fringe field.

TABLE II. Uncertainties for $\Delta P_\mu^\pi \xi$, defined to be the difference of $P_\mu^\pi \xi$ in data minus $P_\mu^\pi \xi$ in simulation. The uncertainties are symmetric except where noted.

| Category | Uncertainty ($\times 10^{-4}$) |
|---|---|
| Target independent | |
| Depolarization in fringe field: | |
|     Alignment of beam and field | +6.4, −1.2 |
|     Transverse field components | +13.9, −0.0 |
|     Multiple scattering | 3.1 |
|     Others | 2.3 |
| Depolarization in stopping material | 3.2 |
| Pion decays in beam line | 1.0 |
| Momentum calibration | 1.5 |
| Chamber response | 2.3 |
| Radiative corrections and $\eta$ | 1.2 |
| Resolution | 1.5 |
| Positron interactions[a] | 0.4 |
| Others | 0.4 |
| Ag target: | |
|     Bremsstrahlung rate | 0.5 |
|     Target thickness and stopping position | 0.6 |
|     Statistical | 4.2 |
| Al target: | |
|     Bremsstrahlung rate | 0.3 |
|     Target thickness and stopping position | 0.8 |
|     Statistical | 3.9 |
| Weighted systematic uncertainty | +16.5, −6.3 |
| Weighted statistical uncertainty | 2.9 |
| Total uncertainty | +16.8, −6.9 |

[a]excluding bremsstrahlung

An uncertainty arises from instabilities in the TEC measurements. The muon beam was measured with the TECs at the beginning and end of each data set, and these measurements should be identical. However, when they are compared, there are variations in the mean position and angle of up to 0.18 cm and 3.0 mrad that were traced to nonreproducibility of the TEC locations when being inserted and removed from the beam line. The beam element stability was verified with continuously monitored measurements of currents, voltages, and temperatures. Variations in the proton beam at the graphite production target were confirmed to have a negligible effect on the muon beam at the TECs. The muon beam itself was demonstrated to be stable by monitoring its mean position in the tracking chambers, and by studying set 72 for which the TECs were in place throughout. Temperature changes

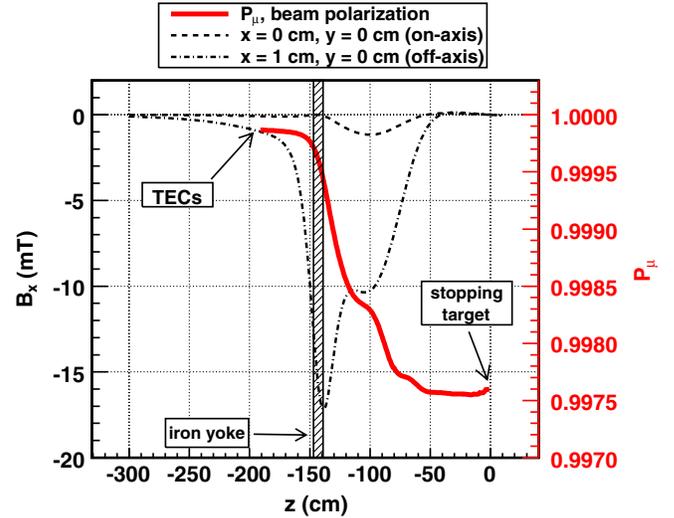

FIG. 9 (color online). The transverse magnetic field component $B_x$ from OPERA (dashed and dash-dotted lines), and the simulated evolution of the muon beam polarization due to the magnetic field (solid line). The on-axis field is nonzero for $z \approx -100$ cm due to the asymmetry of the solenoid's field.

that could have influenced the space-time relationships within the TECs were observed to be too small.

The magnetic field map was aligned in the detector coordinate system with an error of $\pm 0.1$ cm in position and $\pm 1.0$ mrad in angle. The alignment uncertainties for the muon beam measurement and magnetic field map were uncorrelated; an estimate of their combined influence on systematic uncertainties in polarization was obtained by running numerous simulations that sampled random changes in alignment of the TECs and the magnetic field. The sampling simulations resulted in an asymmetric distribution for the polarization at target entry with rms half-widths (with respect to the most probable value) of $(+1.2, -6.4) \times 10^{-4}$. The asymmetric uncertainty arises because the muon beam was well optimized with respect to the solenoidal field so that any misalignment introduced by the sampling simulations tended to lower the beam polarization. Note that the asymmetric uncertainty for $\Delta P_\mu^\pi \xi$ in Table II is opposite to the polarization uncertainty because the latter affects the simulation spectrum, which is subtracted from the data spectrum.

The magnetic field map for the analysis, $(B_x, B_y, B_z)$, was obtained from an OPERA finite element analysis [29]. The longitudinal field components from OPERA ($B_z$) were validated by comparing to Hall probe measurements and were found to have an accuracy of $<0.15$ mT in the region of the DCs, and $<6$ mT within the fringe field region ($z < -60$ cm). This level of error had a negligible effect on the final polarization. However, the transverse field components ($B_x$ and $B_y$) were not measured, and it turned out that the final polarization of the muons was very sensitive to uncertainties in these components within the fringe field region.

There were limitations in the finite element analysis that affected the accuracy of modeling the transverse field components. For example, the magnetic field calculations did not include the reinforcing steel bars in the floor of the experimental hall, and details of the solenoid's coils (position and current density) could only be tuned using the $B_z$ components. Additionally, it is reasonable to expect that the finite element analysis had inaccuracies when reproducing the iron-air interface around the yoke hole ($z \approx -150$ cm), where there are sharp features with length characteristics on a scale of ~0.1 cm. The last beam line quadrupoles Q6 and Q7 were not included in the map used for the analysis, but a separate OPERA study confirmed that the distortion of the solenoid fringe field due to the iron of these magnets had a negligible effect on the beam polarization.

The accuracy of the polarization simulation and the OPERA field map were tested using data sets 72, 76, and 86. As described in Sec. IV, the muon beam for these data sets had significantly reduced polarization compared to the nominal beam. In Table III the reductions in polarization from data and simulation are listed. The simulation underestimates the reduction by $2.0\sigma$ and $0.6\sigma$ for sets 74 compared with 76, and 87 compared with 86, respectively. It overestimates the reduction by $1.0\sigma$ for sets 74 and 72. We presume that the $2.0\sigma$ difference for set 74 compared to 76 has its origins in the simulation of the depolarization or the quality of the OPERA field map. The agreement between data and simulation for the three entries from Table III was quantified as a function of scaled transverse components relative to their OPERA values, using a weighted squared difference statistic similar to chi-squared minimization [27]. The best agreement was found for an increase of 10%. Simulations with these values of transverse field components yielded an average reduction in polarization of $13.9 \times 10^{-4}$. An independent approach of adding fields from three extra on-axis coils to almost completely eliminate the mismatch between the measured $B_z$ component and the OPERA map gave a consistent value for the depolarization. Thus the uncertainty in $\Delta P_\mu^\pi \xi$ from the accuracy of the fringe field is taken to be $(+13.9, -0.0) \times 10^{-4}$.

Another source of uncertainty arose from the correction applied to account for MCS in the TECs using a GEANT simulation (Sec. V). The accuracy of the simulation was tested with data for which an extra 19 $\mu$m Mylar layer had been inserted next to the upstream window of the TEC enclosure, and comparing the increase in MCS from data with a corresponding simulation. The simulation overestimated the increase in the width of the angle distributions by 17%, where the systematic errors on this quantity have not been evaluated. This leads to an uncertainty in the predicted polarization of $\pm 3.1 \times 10^{-4}$. The other fringe field depolarization uncertainties are a limitation in distinguishing signal from noise in the TECs ($\pm 1.7 \times 10^{-4}$) and aging of the TEC sense planes ($\pm 1.5 \times 10^{-4}$). Combined in quadrature, the uncertainty for these two contributions is $\pm 2.3 \times 10^{-4}$.

### B. Depolarization in stopping material

Following muon thermalization inside the stopping target, the polarization underwent time-dependent relaxation that must be quantified. The amount of depolarization was determined with an exponential fit to the data (Fig. 10). Further detail on the analysis that determined the relaxation constant $\lambda$ is given in Appendix A. Preliminary relaxation values were used in our simulation, and the final values are in Table IV. Over the time range used in the analysis ($1.05 < t < 9.00$ $\mu$s), the lifetime-weighted stopping material depolarization was $25 \times 10^{-4}$ for Ag and $39 \times 10^{-4}$ for Al.

Table IV includes consistent results from a subsidiary muon spin relaxation ($\mu^+$SR) experiment [24], which allowed us to establish that no additional fast depolarization takes place in the time range between 10 ns and 1 $\mu$s; this time range could not be used in data from the TWIST detector array due to the time overlap of muon and positron ionization in the drift chambers.

The systematic uncertainty for $\lambda$ originates from a possible analysis bias. Specifically, when we applied the analysis to simulation, the weighted average of $\lambda$ for the Al simulations was consistent with the input value, but for

TABLE III. Polarization difference between two data sets, where the muon beam for one of the sets has sampled a region of the fringe field with increased transverse magnetic field components. The data uncertainties are statistical. The simulation uncertainties are systematic from the possible position/angle misalignments of the muon beam measurement and the accuracy of the magnetic field map.

| Data sets | | Polarization difference ($\times 10^{-4}$) | | | |
|---|---|---|---|---|---|
| | Data | OPERA field | | $B_x$, $B_y$ increased by 10% | |
| | | Simulation prediction | Simulation minus data | Simulation prediction | Simulation minus data |
| 74–76 | $105 \pm 9$ | $56^{+23}_{-18}$ | $-48^{+24}_{-20}$ | $69^{+27}_{-26}$ | $-36^{+28}_{-28}$ |
| 87–86 | $62 \pm 8$ | $47^{+24}_{-16}$ | $-15^{+24}_{-18}$ | $60^{+26}_{-19}$ | $-2^{+27}_{-20}$ |
| 74–72 | $18 \pm 9$ | $28^{+12}_{-5}$ | $+10^{+15}_{-10}$ | $35^{+18}_{-6}$ | $+17^{+20}_{-11}$ |

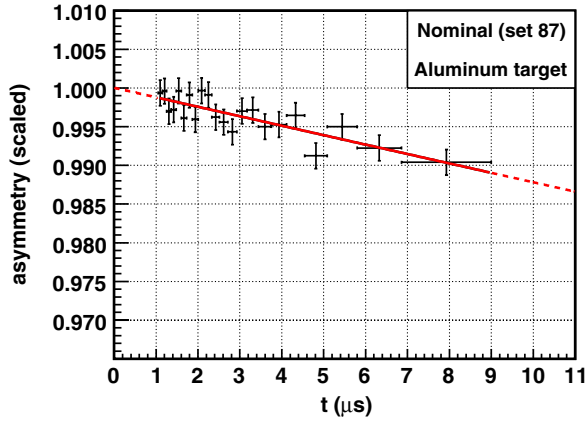

FIG. 10 (color online). Example of an exponential fit to the forward-backward asymmetry from set 87. The vertical scale factor is arbitrary.

the Ag simulations the average was $2.4\sigma$ below the input value. The agreement could not be improved by changing the binning, the asymmetry weighting factors, or the time range of the fits. A systematic uncertainty is assigned to cover the possibility that an analysis bias exists.

The statistical uncertainties in the determination of $\lambda$ introduced $P_\mu^\pi\xi$ uncertainties of $\pm2.4\times10^{-4}$ for both the Ag and Al targets. The systematic uncertainty for $\lambda$ introduced an uncertainty for both targets of $\pm3.0\times10^{-4}$. This was added in quadrature with other small systematic uncertainties in the depolarization from muons that passed the analysis cuts but did not stop in the metal target. These included muons that stopped in the PC 6 chamber gas before reaching the stopping target and were not removed by pulse width discrimination (Fig. 6) ($\pm0.3\times10^{-4}$), muons that passed through the target and stopped in PC 7 but did not produce a signal ($\pm0.9\times10^{-4}$), and muons that scattered back from the target into PC 6 ($\pm0.2\times10^{-4}$).

### C. Uncertainties not related to muon beam polarization

The stopping distribution in the simulation must match the data in order to account for the correlation of muon polarization and range. Specifically, the highest polarization muons have the longest range since their angles with respect to the magnetic field in the tracking region are small. The stopping distributions in data and simulation have small disagreements in position and shape that introduce an uncertainty of $\pm1.0\times10^{-4}$ in $P_\mu^\pi\xi$.

TABLE IV. Results for the relaxation parameter, $\lambda$, from fits of $P_\mu(t) = P_\mu(0)\exp(-\lambda t)$ to the TWIST data, and to a subsidiary $\mu^+$SR experiment [24].

| Experiment | Target | Parameter $\lambda$ (ms$^{-1}$) |
|---|---|---|
| TWIST | Ag | $0.82 \pm 0.08$(stat.) $\pm0.10$(syst.) |
| | Al | $1.28 \pm 0.09$(stat.) $\pm0.10$(syst.) |
| $\mu^+$SR | Ag | $1.0 \pm 0.2$(stat.) $\pm0.2$(syst.) |
| | Al | $1.3 \pm 0.2$(stat.) $\pm0.3$(syst.) |

Two models were applied to correct for a discrepancy in reconstructed momentum of $\sim0.01$ MeV/$c$ at the spectrum end point between data and simulation [(Eqs. (16) and (17)]. This discrepancy was larger than expected from the considered errors, which included the simulation's muon stopping distribution, the thickness of the stopping and detector materials, the scale of the magnetic field map used by the analysis, and the simulation's positron energy loss physics. Without additional information to determine the cause of the energy scale discrepancy, half the difference between the two calibrations amounted to a systematic uncertainty for $P_\mu^\pi\xi$ of $\pm1.4\times10^{-4}$. This was added in quadrature with a contribution from uncertainties in the magnetic field map within the positron tracking region to produce a total uncertainty due to momentum calibration of $\pm1.5\times10^{-4}$.

The following effects were grouped into the chamber response category: differences in the calibration of the DC space-time relationships between data and simulation ($\pm0.9\times10^{-4}$), the shape and position of the cathode foils relative to the wires ($\pm1.3\times10^{-4}$), differences in the upstream and downstream track reconstruction efficiencies in the real detector ($\pm1.4\times10^{-4}$), crosstalk that was not fully removed from the data ($\pm0.5\times10^{-4}$), and uncertainties in determining the wire time offsets ($\pm0.8\times10^{-4}$). These uncorrelated effects were added in quadrature to give a systematic uncertainty in $P_\mu^\pi\xi$ from chamber response of $\pm2.3\times10^{-4}$. Note that the wire time offset precision was significantly improved over our previous measurement of $P_\mu^\pi\xi$ due to the addition of a scintillator at the downstream end of the detector (see Sec. II), which allowed the offsets to be calculated for each data set.

Two pieces of external information were used by the analysis: the muon radiative corrections as approximated by published theoretical calculations, and the muon decay parameter $\eta = (-36 \pm 69) \times 10^{-4}$ [30], which cannot be extracted with as high precision from our decay spectrum. Uncertainties in the external information lead to a total $P_\mu^\pi\xi$ uncertainty of $\pm1.2\times10^{-4}$.

Because the spectrum in the fiducial region was smooth and varied very slowly in momentum and angle, the decay parameters were insensitive to a difference in resolution between data and simulation. Only at the sharp end point could a difference in momentum resolution be detected, and no significant difference was observed. An analysis [12] of sets 73 and 80 where the muons were stopped at the entrance of the detector together with matching simulations yielded a comparison of the resolutions for momentum and angle. The difference between data and simulation results in a conservative uncertainty for $P_\mu^\pi\xi$ of $\pm1.5\times10^{-4}$.

The systematic uncertainty for positron interactions included a contribution due to materials external to the tracking volume that were absent from the simulation (e.g., tracks were not simulated inside the iron yoke), leading to an uncertainty in $P_{\mu}^{\pi}\xi$ of $\pm 0.4 \times 10^{-4}$ due to missing backscattered decay positrons. There were two other contributions to the positron interactions' uncertainty from the simulation of bremsstrahlung and delta-electron production. The rates for these two processes in the simulation were compared to data by selecting positron trajectories that were identified by the reconstruction software as two distinct segments, corresponding to an event where a discrete energy loss or scattering process took place. By comparing the energy difference between the two track segments [12], the bremsstrahlung rate in data for energy loss between 15 and 35 MeV/$c$ was found to be greater than the simulation by $(2.4 \pm 0.4)\%$, leading to target-dependent uncertainties of $\pm 0.5 \times 10^{-4}$ for Ag and $\pm 0.3 \times 10^{-4}$ for Al. The delta-electron rate was measured by selecting tracks that were broken into two segments, with a third segment from an additional negatively charged track that originated from the break point as a delta-electron candidate. The candidates were reconstructed if their momentum exceeded $\sim 6$ MeV/$c$. The simulation matched the data for the yield of delta-electrons at the level of $(-0.7 \pm 0.9)\%$, resulting in a systematic uncertainty of $\pm 0.1 \times 10^{-4}$.

Lastly, the uncertainties marked "Others" in Table II include the effect of errors in the detector length scale and the beam intensity. Uncertainties that were found to be below $0.3 \times 10^{-4}$ are not included in the table, such as the alignment of the DCs (in position and in angle), intensity of beam line positrons, position of the magnetic field in the DC tracking region, and bulging of the DC cathode foils due to gas pressure variations.

## VIII. CORRECTIONS AND RESULTS

The set-by-set results for $\Delta P_{\mu}^{\pi}\xi$ shown in Table V include all set-dependent corrections that are described below; they are averaged over the two momentum calibration approaches, Eqs. (16) and (17). The uncertainties are the quadratic sum of statistical errors from the fits and set-dependent statistical uncertainties from the momentum calibration. The correlation coefficients for the fits are given in Table VI.

There were four corrections made to $\Delta P_{\mu}^{\pi}\xi$ before unblinding the result. The first one corrected for muon depolarization inside the graphite production target, which was not in the simulation. For nominal beam settings, muons were selected from pion decays at an average depth of 16 $\mu$m from the surface of the production target. While exiting the target, MCS changed the muons' momentum vectors but not their spin vectors, so that the spin and momentum were no longer exactly antiparallel. A correction to $\Delta P_{\mu}^{\pi}\xi$ of $+0.9 \times 10^{-4}$ was required for the nominal data sets, $+5.9 \times 10^{-4}$ for a beam momentum of 28.75 MeV/$c$, and $+5.2 \times 10^{-4}$ for a beam momentum of 28.85 MeV/$c$. Within the M13 beam line the muons experienced magnetic fields. Because of the anomalous magnetic moment of the muon, there is a difference in the rotation frequency of the momentum vector and the precession frequency of the spin vector that can change the angle between them. This reduction in polarization is negligible ($P_{\mu}$ changes at the level of $\sim 10^{-8}$).

TABLE V. Difference between $P_{\mu}^{\pi}\xi$ in data and simulation, averaged over the momentum calibration models. There are 2439 degrees of freedom for each fit.

| Set | Target | $\Delta P_{\mu}^{\pi}\xi(\times 10^{-4})$ | Reduced $\chi^2$ |
|-----|--------|-------------------------------------------|------------------|
| 68 | Ag | 90.7 ± 7.6 | 0.975 |
| 70[a] | Ag | 78.7 ± 6.3 | 0.974 |
| 71[a] | Ag | 92.8 ± 6.6 | 0.995 |
| 72[a] | Ag | 90.5 ± 6.4 | 1.029 |
| 74 | Ag | 84.1 ± 7.5 | 1.002 |
| 75 | Ag | 84.6 ± 6.4 | 1.006 |
| 76[a] | Ag | 33.3 ± 7.0 | 0.995 |
| 83 | Al | 82.2 ± 6.6 | 0.988 |
| 84 | Al | 70.5 ± 6.9 | 1.030 |
| 86[a] | Al | 53.9 ± 6.2 | 0.994 |
| 87 | Al | 83.6 ± 6.7 | 0.988 |
| 91 | Al | 83.3 ± 13.0 | 1.054 |
| 92 | Al | 74.9 ± 11.2 | 1.015 |
| 93 | Al | 63.7 ± 9.2 | 1.030 |

[a]used only for systematics studies

The second correction accounted for the differences between preliminary and final $\lambda$ values. These were $+2.9 \times 10^{-4}$ for the Ag target and $+2.4 \times 10^{-4}$ for Al. Two additional corrections arose because for each data set we generated a simulation with more events than the data, by a factor of between 1.8 and 4.0. This imbalance in statistics introduced a small bias [12] when fitting data to simulation spectra, requiring a set-*independent* correction of $-0.5 \times 10^{-4}$. The statistics imbalance also introduced a bias in the momentum calibration, requiring a set-*dependent* correction of between $+1.3 \times 10^{-4}$ and $+2.3 \times 10^{-4}$.

TABLE VI. Correlation coefficients from (a) the fits of data to simulation, (b) average correlations for the systematic uncertainties common to $\rho$, $\delta$, and $P_{\mu}^{\pi}\xi$, and (c) correlations for the final result, which are used for the calculation of the errors on $P_{\mu}^{\pi}\xi\delta/\rho$. The "+" and "−" labels indicate correlations for the asymmetric errors.

| Category | $\rho\delta$ | $\rho\xi$ | $\delta\xi$ |
|----------|--------------|-----------|-------------|
| (a) | 0.190 | 0.206 | −0.719 |
| (b) | 0.677 | | |
| (b)+ | | −0.072 | −0.162 |
| (b)− | | −0.190 | −0.429 |
| (c) | 0.532 | | |
| (c)+ | | −0.001 | −0.359 |
| (c)− | | 0.051 | −0.177 |

Our final $P_\mu^\pi \xi$ result did not use sets 72, 76, and 86 since their systematic uncertainties for the polarization were not understood sufficiently well. Sets 70 and 71 were also excluded because of unevaluated fringe-field uncertainties; the fringe field was not measured at 1.96 T, 2.04 T, and the maps were not calculated with OPERA.

After revealing the hidden parameters, the central value for the decay parameter $P_\mu^\pi \xi$ was 1.000 84 with errors of $\pm 0.000\,35(\text{stat.}) {}^{+0.001\,65}_{-0.000\,63}(\text{syst.})$. However, this is not the final result of our analysis. Since the TWIST experiment simultaneously also measured $\rho$ and $\delta$, we could compute the product $P_\mu^\pi \xi \delta / \rho$ including correlations (see Table VI). If Eqs. (2)–(4) are evaluated without radiative corrections, at the maximum energy ($x = 1$) and for both extremes of angle ($\cos\theta = \pm 1$), an asymmetry can be formed that is given by this product. Being an asymmetry, it is constrained to be less than or equal to one. Not only did we find that our decay parameter values combined to give $P_\mu^\pi \xi \delta / \rho = 1.001\,92 {}^{+0.001\,67}_{-0.000\,66}$, but more significantly, the value of $P_\mu^\pi \xi \delta / \rho$ for the Ag data was higher than that for Al by $3.8\sigma$. As a consequence, we reviewed each category of systematic uncertainties to look for possible mistakes and searched for effects that might have been overlooked in the blind analysis.

Several factors decreased the likelihood of finding a simple problem in the analysis. The quantity $P_\mu^\pi \xi \delta / \rho$ is very insensitive to most categories of systematic uncertainty. The correlation coefficients (Table VI) show that changes in $\rho$ and $\delta$ typically have the same sign and similar magnitudes, which is reflected in the correlation coefficient for the common systematics. The sensitivity of $\xi$ to non-polarization systematics is typically small. A reduction of the measured $P_\mu^\pi \xi \delta / \rho$ by as much as 0.0019 due to an error in simulating the change of polarization would require an *increase* in the simulated value of $P_\mu$ by the same amount. In the simulation, the difference of $P_\mu$ from 1.0 is dominated by fringe-field effects of about 0.0025 (Fig. 9). We consider it unlikely that this difference could be overestimated by several times the entire positive systematic uncertainty of $\Delta P_\mu^\pi \xi$ (Table II), as would be required to explain our value of $P_\mu^\pi \xi \delta / \rho$.

No significant mistakes were found in the estimates of the systematic uncertainties previously considered. However, we did find that two corrections had been missed. A correction of $0.3 \times 10^{-4}$ for muon radiative decay was added for the Ag data due to photons converting in the target. This correction was negligible for Al. A second correction was made for each data set to account for effects in the analysis of a difference between the mean muon stopping position for data and simulation. We had expected the momentum calibration to remove any such effects.

However, the algorithms of Eqs. (16) and (17) did not compensate for the difference due to bremsstrahlung, which distorts the spectrum in a specific pattern at large energy losses. The magnitude of the difference in muon stopping position between data and simulation averaged $0.7\ \mu m$ for Ag and $1.5\ \mu m$ for Al [12]. This was determined for each data set from the tails of the stopping distribution, i.e., muon ionization tracks that terminated in chambers other than PC 6 and thus were muons stopping outside the target.

We also concluded that the uncertainties for the two targets were sufficiently different for $\rho$ and $\delta$ to merit dividing the systematic uncertainties into common and target-dependent categories. The target-independent systematics are unchanged from the blind analysis. Separate uncertainties for bremsstrahlung were computed, and an additional sensitivity to the muon stopping position in each target was added.

The final result includes the same data sets as had been chosen prior to revealing the hidden parameters, so our value of $P_\mu^\pi \xi$ is still based on sets excluding 70, 71, 72, 76, and 86. The averages were computed separately for the Ag and Al target data, using weights corresponding to statistical uncertainties. The corrections for radiative decay and unequal statistics as described above were then applied. Also, because the results for the $\rho$ and $\delta$ parameters did *not* exclude sets 70, 71, 72, 76, and 86, and because there is a large correlation between $\delta$ and $\xi$ in the fit (see Table VI), the central value and statistical uncertainty for $P_\mu^\pi \xi$ were adjusted to reflect the additional information from the correlation. The statistical error on the determination of $\lambda$ is included separately for the two targets. The weights for combining the results from the two targets include the statistical and target-dependent systematic uncertainties.

The foregoing procedures reduced the difference between targets for $P_\mu^\pi \xi \delta / \rho$ to $\sim 1\sigma$, and resulted in our final value for $P_\mu^\pi \xi$:

$$P_\mu^\pi \xi = 1.000\,84 \pm 0.000\,29(\text{stat.}) {}^{+0.000\,65}_{-0.000\,63}(\text{syst.}). \tag{18}$$

Note that the systematic uncertainties are the same as those of the blind result, and the central value has not changed due to small offsetting effects. Our result is consistent with the standard model values of $P_\mu^\pi = \xi = 1$ and represents an improvement in precision over the pre-TWIST direct measurement [31] by a factor of 7. The new result, which supersedes our earlier result [8], is compared to previous experimental values in Fig. 11.

## IX. LIMITS IN LEFT-RIGHT SYMMETRIC MODELS

Figure 12 shows the allowed regions for the manifest and generalized LRS models [3], derived using only the new value of $P_\mu^\pi \xi$ in addition to contours that use the $P_\mu^\pi \xi$, $\rho$ and $\delta$ results [11].

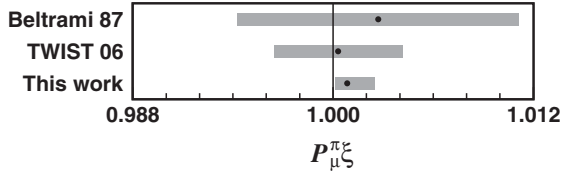

FIG. 11. Summary of the result of this experiment together with previous $P_\mu^\pi \xi$ results [8,31].

In the manifest LRS model, our result restricts the mixing angle to $-0.020 < \zeta < +0.017$ (90% C.L.) compared to the pre-TWIST limit of $-0.064 < \zeta < +0.053$. While our limit is the most restrictive from muon decay, for the manifest LRS the limit from CKM unitarity [32] is $|\zeta| < 0.0005$ (90% C.L.). The lower mass limit from $P_\mu^\pi \xi$ measurements is increased from $m_2 \geq 318$ GeV/$c^2$ to

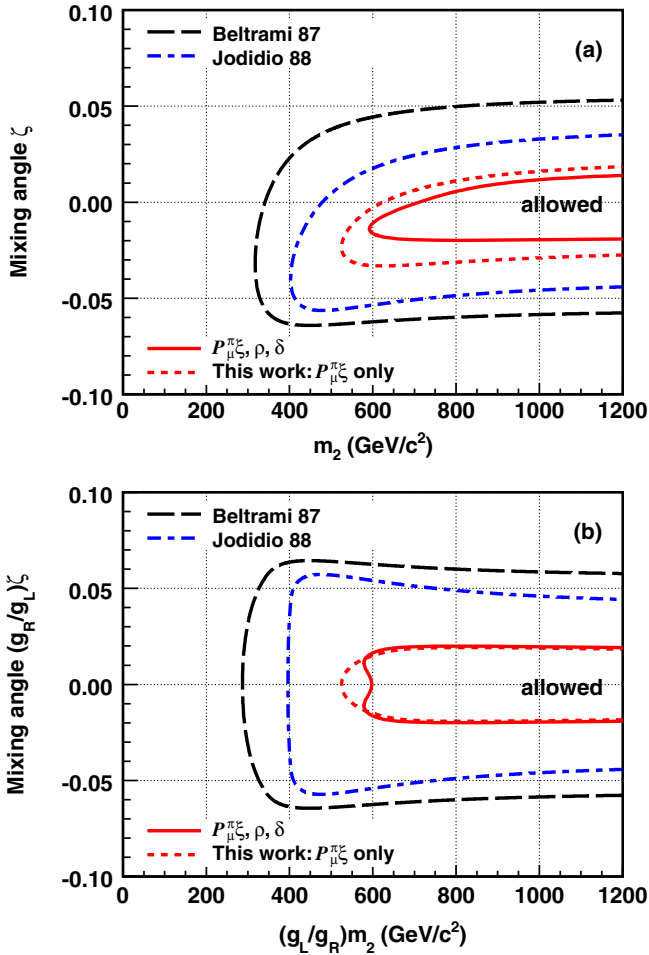

FIG. 12 (color online). Allowed regions (90% confidence) for the $W_2$ mass and mixing angle in (a) manifest and (b) generalized LRS models. We include new limits using only $P_\mu^\pi \xi$, and $P_\mu^\pi \xi$ combined with $\rho$ and $\delta$ from [11].

$\geq 592$ GeV/$c^2$ (90% C.L.) for the manifest LRS model. This new mass limit from muon decay is lower than the result of a direct search for the $W'$, a heavier counterpart to the $W$ with SM couplings, where $m_{W'} > 1.36$ TeV/$c^2$ (95% C.L.) [33,34].

In the generalized LRS model, our new result limits the mixing angle to $|(g_R/g_L)\zeta| < 0.020$ (90% C.L.), compared to the pre-TWIST limit of $|(g_R/g_L)\zeta| < 0.066$. The lower mass limit for $W_2$ $((g_L/g_R)m_2)$ has been increased from 400 GeV/$c^2$ to 578 GeV/$c^2$. Note that these limits make no assumptions about a possible difference of CKM couplings in the left- and right-handed sectors; any values are allowed. Neither is any assumption necessary regarding the possible presence of *CP*-violating phases in the LRS model. Compared with other methods of placing bounds for mass and mixing angle in the generalized LRS model [35], this is a distinct and unique advantage of muon decay where the strong interaction effects are absent in lowest order, except for a small contribution from the polarization of muons resulting from pion decay. As such, our limits represent the most restrictive available on mass and mixing in the generalized LRS model, subject only to the qualification of a light right-handed neutrino.

## X. CONCLUSION

The task of improving the polarization parameter $P_\mu^\pi \xi$ in polarized muon decay was met by the construction of a dedicated spectrometer of high precision and reliability, the use of a well-characterized beam, and an analysis procedure that was designed to control systematic uncertainties as much as possible. While it was understood that systematic issues would dominate precision for all measured muon decay parameters, the challenges presented for $P_\mu^\pi \xi$ were quite different from those of $\rho$ or $\delta$. The result could not have been achieved without several cycles of measurement and analysis that stretched over nearly a decade. The early phases of the program showed where the main difficulties would be; while they were mostly anticipated, the early experiences allowed us to test methods to improve beam quality and characterization as well as methods of evaluation of systematic uncertainties. The result shown in Eq. (18) quantifies parity violation in muon decay with a precision that is a factor of 7 improvement over prior experiments, setting new limits on possible mass and mixing angle combinations in a generalized (nonmanifest) left-right symmetric model.

## ACKNOWLEDGMENTS

We thank all early TWIST collaborators and students for their substantial contributions, in particular, B. Jamieson whose thesis work described the first $P_\mu^\pi \xi$ result from TWIST. We also thank C. Ballard, M. Goyette, S. Chan, and the TRIUMF cyclotron operations, beam lines, and support personnel. Computing resources were provided by WestGrid and Compute/Calcul Canada. This work was supported in part by the Natural Sciences and Engineering Research Council and the National Research Council of Canada, the Russian Ministry of Science, and the U.S. Department of Energy.

## APPENDIX A: TIME-DEPENDENT DEPOLARIZATION ANALYSIS

It is necessary to determine the depolarization rate $\lambda$ in the stopping material in order to account for the change in $P_\mu$ between the time when the muon stops and when it decays (see Sec. VII B). In our previous measurement [8], this was accomplished by calculating the time dependence of the forward-backward asymmetry:

$$\epsilon(t) = \frac{N_F(t) - N_B(t)}{N_F(t) + N_B(t)}, \qquad (A1)$$

where $N_F$ ($N_B$) is the total yield within the fiducial region in the forward (backward) direction. This treated all events as equivalent, independent of the sign or magnitude of their expected asymmetry, so it did not provide the most efficient use of the available statistics.

For the present measurement, Eq. (A1) was replaced with a weighted asymmetry calculation. If SM values are assumed for $\rho$, $\delta$, and $\xi$, and radiative corrections and the positron mass are neglected, then Eqs. (2)–(4) reduce to

$$\frac{d^2\Gamma}{dx\,d\cos\theta} \propto x^2(3 - 2x)[1 + P_\mu A(x)\cos\theta], \qquad (A2)$$

where

$$A(x) = \frac{2x - 1}{3 - 2x}. \qquad (A3)$$

After exploring a number of possible weight functions, we chose to replace Eq. (A1) by

$$\epsilon(t) = \frac{\sum w_i N_i(t)}{\sum |w_i| N_i(t)}, \qquad (A4)$$

where the sum is over the yield in the forward and backward directions, and

$$w(x, \cos\theta) = A(x)\cos\theta|A(x)\cos\theta|. \qquad (A5)$$

This assigns very small weights to events with low or intermediate momenta so they make a negligible contribution to the statistical precision of this asymmetry. Thus, we only considered events with $p > 31$ MeV/$c$ when calculating Eq. (A4). Figure 10 illustrates the determination of $\lambda$ using Eq. (A4).